\documentclass[12pt]{article}
\usepackage{epsf}
\newcommand{\be} {\begin{equation}}
\newcommand{\ee} {\end{equation}}
\newcommand{\bdm} {\begin{displaymath}}
\newcommand{\edm} {\end{displaymath}}
\newcommand{\bc} {\begin{center}}
\newcommand{\ec} {\end{center}}
\newcommand{\beqa} {\begin{eqnarray}}
\newcommand{\eeqa} {\end{eqnarray}}

\newcommand{\veE}{\mbox{\boldmath${\rm E}$}}

\newcommand{\vex}{\mbox{\boldmath${\rm x}$}}

\begin{document}
\bc
{\bf \LARGE Finite temperature correlation functions}
\ec
\vspace{.5cm}
\bc
{\large  D.S.Kuzmenko\footnote{e-mail: kuzmenko@heron.itep.ru},
Yu.A.Simonov\footnote{e-mail: simonov@heron.itep.ru}}
\ec
\bc
{\large \it Institute of Theoretical and Experimental Physics,\\
 Moscow, Russia}
\ec

\begin{abstract}
Lattice measurements of the Pisa group (A.Di
 Giacomo et al., hep-lat/9603018) are analyzed  numerically and
 parameters of correlation functions
are extracted from the data -- both below and above deconfinement
temperature $T_c$. Gluon condensate is found for six temperatures
in the interval 0.956 $T_c$ -- 1.131 $T_c$ and field distributions in
deconfined phase are obtained.
\end{abstract}

\section{Introduction}

Study of field distributions around static quarks has a long
history. The information obtained both analytically and on the
lattice has an important meaning in
several respects. Firstly, it demonstrates clearly the appearance
of the QCD string between the charges in the confining phase and
the detailed contents of the fields in this string; e.g., the
string consists mainly of longitudinal color-electric field.  Secondly,
since the string enters in many dynamical quantities, such as static
interquark potential, spin-dependent forces for heavy quarkonia etc., one
can easily compute these quantities  from the string field distributions.

Finally, and this is the purpose of the present paper, the
phenomenon of deconfinement is not fully understood from the point
of view of the string and field distributions.

What actually happens when temperature $T$ exceeds the critical
value -- the string disappears or distribution
 of fields drastically change, so that to compensate the
 string tension?

 Happily there exist numerical lattice measurements of field
 correlators near the critical temperature $T_c$, made by the Pisa
 group [1], where both electric and magnetic correlators are found with
 good accuracy. These data clearly  demonstrate the strong
 suppression of color-electric component above $T_c$ and persistence
 of color-magnetic components.

 The purpose of the present paper is twofold. First, we reanalyze
 the data of \cite{1} in terms of correlators $D^\mathrm{E,B}$ and
 $D_1^\mathrm{E,B}$ which are better understood from the point of view of
 perturbative and nonperturbative contributions [2]. There
was e.g. shown that the string tension $\sigma$ ("conventional" electric
or "spatial" magnetic) is expressed directly through the
 $D^\mathrm{E,B}$ but not the $D_1^\mathrm{E,B}$.  Given the simple
assumption of behavior $D^\mathrm{E,B}(x)$ and $D_1^\mathrm{E,B}(x)$ in all
region of distances $x$ between correlation points, we obtain the behavior
of gluon condensate near the $T_c$ (which is defined by $D^\mathrm{E,B}(0)$
and $D_1^\mathrm{E,B}(0)$).
 Second, using the obtained $D^\mathrm{E,B}(x)$ and $D_1^\mathrm{E,B}(x)$, we calculate
the color field distributions around static quarks in the deconfined phase.
 We perform these calculations using the connected probe [3]
in the framework of Field Correlator Method (FCM) [4,5].
We study in detail two possible from the deconfinement point of view
  regimes, corresponding to two forms of $D(x)$, extracted from
 lattice data [1]. In the first one the string disappears  and in the second
 (less appealing physically, but more supported by the lattice data)
 the string becomes a coaxial cable with the empty core.

 The paper is organized as follows.

In Section 2 a short description is given of our fitting procedure
of magnetic correlators at any $T$ and the electric correlators at $T<T_c$,
 while in Section 3 more prolonged one is given for
electric correlators at $T>T_c$. The Section 4 is devoted
to the behaviour of the gluonic condensate below and above $T_c$.
In Section 5 the detailed equations for the field distributions around
deconfined quark and antiquark are given and illustrated
graphically.
In the concluding section results of the paper are summarized and
discussed.

\section{Fitting data with nonzero string tension}

In this section we consider magnetic correlators in all
temperature region and electric correlators at $T<T_c$ since both
of them  produce (e.g., spatial) nonzero string tension.
We fit the data [1] using the  method of least squares [6].
To begin with, we express the functions $D_{||}(x),~D_{\bot}(x)$
through the $D(x),~D_1(x)$ and represent the lasts as  sums
of nonperturbative (NP) and perturbative (P) (diverging at zero)
contributions. This is the procedure used by
A.Di Giacomo et al.[7] when analyzing the correlation functions at
 zero temperature. The difference is that we should distinguish
electric and magnetic correlators due to the fact that the finite
temperature theory has only the $O(3)\times O(1)$ symmetry.

 We fit distributions of
 \be
 D_{||}^{\mathrm{E,B}}(x)=D^{\mathrm{E,B}}(x)+D_1^{\mathrm{E,B}}(x)
+\frac{x}{2}\frac{\partial D_1^{\mathrm{E,B}}(x)}{\partial x}
 \label{1}
 \ee
 and
 \be
 D_\bot^{\mathrm{E,B}}(x)=D^{\mathrm{E,B}}(x)+D_1^{\mathrm{E,B}}(x)
 \label{2}
 \ee
 electric and magnetic correlation functions in the range from 0.4 to 1 fm.
  All points are measured at $x_4=0$.
 In this section we parameterize the functions as follows:
 \be
 D(x)=A\exp (-x/\lambda_A)+\frac{a}{x^4}\exp (-x/\lambda_a),
 \label{3}
 \ee
 \be
 D_1(x)=B\exp (-x/\lambda_B)+\frac{b}{x^4}\exp (-x/\lambda_b).
 \label{4}
 \ee

Fitting the  magnetic functions (which are close to exponentials), we set in (3),(4)
 $\lambda_a=\lambda_b=\lambda_B$, thus using $k=6$ fitting parameters. Given $N=12$
number of data, we have as a result   $n=N-k=6$ number of degrees
of freedom. The results are shown in Table 1.
 One-standard-deviation errors $\Delta\alpha_i$ are determined from the equation
$\chi^2(\alpha_1,...,\alpha_i+\Delta\alpha_i,...,\alpha_k)=
\chi^2(\alpha_1,...,\alpha_i,...,\alpha_k)+
 \Delta\chi^2_k$, where $\alpha_i$ are fitted parameters and
  $\Delta\chi^2_1=1$, $\Delta\chi^2_4=4.75$,
 $\Delta\chi^2_6=7$  [6].

Preliminary fitting of electric data at $T=0.956T_c$ with parameters of
(3),(4) gave unacceptable big $\chi^2$ due to the end  points of $D^\mathrm{E}_{||}$
and $D^\mathrm{E}_{\bot}$;
 $\lambda_A,~\lambda_a$, and $\lambda_b$ were found close
to each other. Therefore we had removed mentioned points and set
$\lambda_A=\lambda_a=\lambda_b$ to get the reasonable fit (Table 2).

Preliminary fitting of electric data at $T=0.978T_c$ have shown  that
$\chi^2$  is also too big  and besides  that $\lambda_A$ is
 close to $\lambda_b$ and  $\lambda_B$ to $\lambda_a$. Therefore we set
$\lambda_b=\lambda_A$ and $\lambda_a=\lambda_B$. To improve
$\chi^2$, we enlarge two times the error  of the last point of
$D_{||}^E$ (Fig. 1). One can make sure from Figure 1 that this point is
largely off the exponential curve, which may be connected to the
lattice size effects. The results of fitting are shown in Table 2.

\section{Fitting electric data in the deconfinement region}

 $D^\mathrm{E}_{||}$ above $T_c$ has a drop that is presumably related
with the deconfinement transition, when the string tension of area law
asymptotics of Wilson loop for static quarks disappears. In gaussian
approximation of FCM [5] we get \be
\sigma=\frac1 2 \int dx_1dx_4 D^{\mathrm{E,NP}}(x_1,x_4 )=0.
\label{5}
\ee

 In this region we should alter the form of $D$ (3) to justify (5).
 It is naturally to set
\be
D^{\mathrm{NP}}(x)\equiv 0.
\label{6}
\ee
(We omit subscript "E"  here and in what follows in this section.) As we
shall see below, this form assures reasonably well fit of data.
Let us propose another form of $D$ to ensure good fitting of data.
We suppose now more constrained $O(4)$ symmetry: $D(\vex^2,x_4^2)=
D(\vex^2+x_4^2)\equiv D(x^2)$ to get
\be
   \sigma=\frac{\pi}{2}\int_0^{\infty} dx^2 D^{\mathrm{NP}}(x^2)=
     \frac{\pi}{2}\int_0^{\infty} dx^2 D^{\mathrm{NP}}_{||}(x^2)=0.
\label{7}
\ee
To better reproduce the mentioned drop of data, we set
$D^{\mathrm{NP}}_{||}(x)\equiv 0$, i.e.,
\be
D^{\mathrm{NP}}(x)=-D_1^{\mathrm{NP}}(x)-\frac{x}{2}
\frac{\partial D_1^{\mathrm{NP}}(x)}{\partial x},
\label{8}
\ee
leaving meanwhile $D_1$ in form (4) intact.

So far we have adopted two forms of $D(x)$:
 \be
                 D(x)=\frac{a}{x^4}\exp (-x/\lambda_a);
\label{9}
\ee
\be
 D(x)=B(\frac{x}{2\lambda_B}-1)\exp (-x/\lambda_B)-
               \frac{a}{x^4}\exp (-x/\lambda_a).
\label{10}
\ee
We fit data on $D_{||}$ and $D_{\bot}$ at $T=1.011T_c$ in two
ways, using (a): functions (9),(4) and (b): (10),(4). Having in
mind  the  small number of data we set $\lambda_a= \lambda_b=
\lambda_B$ to have $k=4$. Given $N=9$, we obtain $n=5$ (Table 3).
We see from the Table 3 the somewhat reasonable $\chi^2/n=1.7$
in case (a) and excellent $\chi^2/n=1.05$ in case (b).

At higher temperatures there are only two measurements of
$D_{||}$, with  values significantly less even than the errors of
corresponding points of $D_{\bot}$. This circumstance allows us to
subdivide the fitting procedure in two stages. At the first stage
we fit the difference $D_\bot-D_{||}$ (cf. (1),(2)) by
$-x^2\partial D_1/\partial x^2$ (cf. (4)),
 and extract parameteres ($B, \lambda_B, b, \lambda_b$).
At the second stage we fit $D_\bot$,
reproducing it (in two ways, due to two cases of $D$) by the sum
of  $D$ and $D_1$ with parameters extracted from the first fit;
$D$ is taken in forms (9)
and (10) with $\lambda_a=\lambda_b$ in both cases, and  only
parameter $a$ is  allowed to vary (Table 4, Fig. 2).
One-standard-deviation errors are determined as described above,
with $k=4$ at the first stage and $k=1$ at the second stage of
fitting. In the Table 4 the second--stage--fitting results are separated
by horizontal line; $\chi_{1,2}^2$ refer to the first and second stages
correspondingly.

As $D_\bot-D_{||}\approx D_\bot$, at the first stage we actually fit the
data by
\be
    D_{\bot}(x)=\frac{B x}{2\lambda_B}\exp (-x/\lambda_B)+
               \frac{2 b}{x^4}(1+\frac{x}{4\lambda_b})\exp (-x/\lambda_b)
\label{11}
\ee
and at the second stage in case (a) by
\be
    D_{\bot}(x)=B \exp (-x/\lambda_B)+
               \frac{a+b}{x^4}\exp (-x/\lambda_b)
\label{12}
\ee
and in case (b) by
\be
    D_{\bot}(x)=\frac{B x}{2\lambda_B}\exp (-x/\lambda_B)+
                 \frac{a+b}{x^4}\exp (-x/\lambda_b),
\label{13} \ee
 with all parameters of (12),(13) except $a$ fixed by the
first stage. One could see that (13)  well reproduces (11)
at $a=b$ and any $B$ and $\lambda_B$, for $x\ll 4\lambda_b$,
i.e., in all measured region $0.4$ fm$<x<1$ fm (see Table 4).

\section{Temperature dependence of the gluon condensate}

 The gluon condensate is defined as
\be
G_2=\frac{\alpha_s}{\pi}\langle F_{\mu\nu}^a  F_{\mu\nu}^a \rangle,
\label{14}
\ee
where $\alpha_s$ is the strong coupling constant,
 $F_{\mu\nu}^a$ are the gauge field strengths taken at the point
$x=0$ and the averaging is performed over all vacuum configurations. At
zero temperature the FCM reads $$     \frac{g^2}{N_c} \langle
F_{\rho\sigma}^a(x')(T^a)^{\alpha}_{\beta} \Phi^{\beta}_{\gamma}(x',x)
F_{\mu\nu}^b(x)(T^b)^{\gamma}_{\delta}
\Phi^{\delta}_{\alpha}(x,x')\rangle= $$
$$
 (\delta_{\rho\mu}\delta_{\sigma\nu}-
  \delta_{\rho\nu}\delta_{\sigma\mu})(D(h^2)+D_1(h^2))+
$$
\be
   (h_{\mu}h_{\rho}\delta_{\nu\sigma}-h_{\mu}h_{\sigma}\delta_{\nu\rho}-
    h_{\rho}h_{\nu}\delta_{\mu\sigma}+h_{\nu}h_{\sigma}\delta_{\mu\rho})
    \frac{\partial D_1(h^2)}{\partial h^2},
\label{15}
\ee
where $h\equiv x-x'$; ~$\alpha$,...,$  \delta$  are
color indices. At $x=x'=0$ one uses in (15)
$\mathrm{tr}T^aT^b=\frac1 2 \delta^{ab}$ and gets
\be
    \frac{g^2}{2N_c} \langle F_{\mu\nu}^a  F_{\mu\nu}^a \rangle =
  12(D(0)+D_1(0)).
\label{16}
 \ee
 In what follows we shall  use the  zero temperature
lattice results  [7]:
 \be
D^{\mathrm{NP}}(0)=3.3\times10^8
\Lambda^4_L=129~\mathrm{fm}^{-4},~
D_1^{\mathrm{NP}}(0)=0.7\times10^8 \Lambda^4_L=27~\mathrm{fm}^{-4},
\label{17}
 \ee
where $\Lambda_L=0.025$ fm$^{-1}$ is the fundamental constant of QCD in
lattice renormalization scheme; its value is extracted from the string
tension.

   At finite
temperature one derives from (15)
\be
     \frac{g^2}{2N_c} \langle E_i^a(0)  E_i^a(0) \rangle =
  3(D^\mathrm{E}(0)+D_1^\mathrm{E}(0)),
\label{18}
\ee
 \be
     \frac{g^2}{2N_c} \langle B_i^a(0)  B_i^a(0) \rangle =
  3(D^\mathrm{B}(0)+D_1^\mathrm{B}(0)),
\label{19}
\ee
  where $E_i\equiv F_{0i}$ and $B_i\equiv \frac1 2 \epsilon_{ilm}F_{lm}$.
Note that at finite temperature $D^\mathrm{E}$ and $D^\mathrm{B}$ acquire
subscripts for  symmetry breaking
$O(4)\longrightarrow O(3)\times O(1)$ reason. We have to distinguish
electric and magnetic contributions to condensate:
 \be
    \langle F_{\mu\nu}^a  F_{\mu\nu}^a \rangle = \langle F^2
\rangle_{\mathrm{el}}+   \langle F^2 \rangle_{\mathrm{magn}},
\label{20}
\ee
where
\be
   \langle F^2 \rangle_{\mathrm{el}}\equiv \langle F_{0i}^a  F_{0i}^a
\rangle +\langle F_{i0}^a  F_{i0}^a \rangle= 2 \langle E_i^a E_i^a \rangle,
\label{21}
\ee
\be
  \langle F^2 \rangle_{\mathrm{magn}}\equiv \frac1 4
\epsilon_{ijk}\epsilon_{ijk}   \langle F_{jk} F_{jk} \rangle + \frac1 4
\epsilon_{ijk}\epsilon_{ijk}   \langle F_{kj} F_{kj} \rangle = 2 \langle
B_i^a B_i^a \rangle \label{22}
\ee
Substituting (16), (18)--(22) into (14), one obtains that normalized gluonic condensate is
\be
\frac{G_2(T)}{G^0_2}=\frac{D^\mathrm{E,NP}(0)+D_1^\mathrm{E,NP}(0)+D^\mathrm
{ B,NP } ( 0 ) +D_1^\mathrm{B,NP}(0) }{2(D(0)+ D_1(0))},
 \label{23}
 \ee
where $G_2(T)$ is the gluon condensate at temperature $T$; $G^0_2\equiv
G_2(0)$.

According to our fitting, magnetic condensate  is
$D^\mathrm{B,NP}(0) +D_1^\mathrm{B,NP}(0)=A+B$, with $A,B$ taken from
Table 1.  Electric condensate at  temperature $T<T_c$ is
$D^\mathrm{E,NP}(0)+D_1^\mathrm{E,NP}(0)=A+B$, with $A,B$ taken from Table
2. At $T>T_c$ in case (a) $D^\mathrm{E,NP}(0)=0$,
$D_1^\mathrm{E,NP}(0)=B$, with $B$ taken from Tables 3,4, and in case (b)
$D^\mathrm{E,NP}(0)+D_1^\mathrm{E,NP}(0)=0$.

The behavior of the  condensate and that of its electric and magnetic
constituents with temperature are shown in Tables 5,6. Data
on the whole condensate from these Tables are plotted in Fig.
3. We see that at $T<T_c$ the value of the condensate is close to its
zero temperature value.
At $T>T_c$ in the case (a) there is a fast growth of condensate.
We will discuss its physical meaning in the concluding section.
In the case (b) the condensate value is about half of its
zero temperature value.

\section{Field distributions around deconfined quarks}

In this section we consider NP part of gluodynamical field
generated by static $Q \bar Q$ sources, using the connected
probe [3,8]. There was shown that the only nonzero components in this system
are longitudinal (along quark axis) and transverse electric fields
$E_1(x_1,x_2)$ and $E_2(x_1,x_2)$, where $x_1$ is coordinate along
quark axis and $x_2$ -- distance to the axis.

In case (a), when $D^{\mathrm{NP}}\equiv 0$,  NP part of $D_1$ is $D_1=B\exp(-x/\lambda)$,
where $x=\sqrt{x_1^2+x_2^2+x_4^2}$ (due to axial symmetry we may set $x_3\equiv 0$),
 $\lambda$ means $\lambda_B$.
Here and in what follows we omit subscript "E". From the equations of [8]
$$
    \langle E_1(x_1,x_2) \rangle_{Q \bar Q}^{(a)} =\int_0^R
dx_1'\int_{-\infty}^{\infty}dx_4'       \left(
D_1(h^2)+(h_1^2+h_4^2)\frac{\partial D_1(h^2)}{\partial h^2} \right)=
$$
\be
     B(x_1\sqrt{x_1^2+x_2^2}K_1(\sqrt{x_1^2+x_2^2}/\lambda)-
(x_1-R)\sqrt{(x_1-R)^2+x_2^2}K_1(\sqrt{(x_1-R)^2+x_2^2}/\lambda)),
\label{24}
\ee
$$
        \langle E_2(x_1,x_2)\rangle_{Q \bar Q}=\int_0^R dx_1'\int_{-\infty}^{\infty}dx_4'
        h_1h_2\frac{\partial D_1(h^2)}{\partial h^2}=
$$
\be
        B x_2 (\sqrt{x_1^2+x_2^2}K_1(\sqrt{x_1^2+x_2^2}/\lambda)-
        \sqrt{(x_1-R)^2+x_2^2}K_1(\sqrt{(x_1-R)^2+x_2^2}/\lambda)),
\label{25}
\ee
where $K_1$ is McDonald function. The total field is
\be
       \langle \veE \rangle^2 = \langle E_1 \rangle^2+\langle E_2 \rangle^2.
\label{26}
 \ee
 In Fig. 4 we plot $\langle \veE(x_1,x_2) \rangle^2$ distribution with
parameters  corresponding to the case
$T=1.070T_c$ and $R=2$ fm. We observe two
"volcanoes"  with quarks hidden in their bottoms.
These two spherically symmetrical in coordinate space distributions are
defined as
\be
   E(r)=Br^2K_1(r/\lambda),
\label{27}
\ee
where $r$ is a distance from quark or antiquark. The field at quark
and antiquark positions is zero and linearly rises at small $r$. Maximal
value of field is
\be
E^{\max}= E(1.33\lambda)=0.63B\lambda^2.
\label{28}
\ee

In Fig. 5 the vector field distribution $\langle \veE(x_1,x_2) \rangle$ is
shown in the vicinity of the quark.

In the case (b) the transverse part of the field,  $E_2$, remains the same
(25). Let us calculate using (8) the longitudinal part of the field, $E_1$:
$$
    \langle E_1(x_1,x_2) \rangle_{Q \bar Q}^{(b)} =
\int_0^R dx_1'\int_{-\infty}^{\infty}dx_4'
\left( D(h^2)+D_1(h^2)+(h_1^2+h_4^2)\frac{\partial D_1(h^2)}{\partial h^2}
\right)=
$$
$$
     =\int_{x_1-R}^{x_1}dh_1\int_{-\infty}^{\infty}dh_4(-h_2^2)
      \frac{\partial D_1}{\partial h^2}=
     \int_{x_1-R}^{x_1}dh_1\int_0^{\infty}dh_4 \frac{h_2^2}{h\lambda}D_1=
$$
\be
  =\frac{B x_2^2}{\lambda}\int_{x_1-R}^{x_1}dh_1
K_0(\sqrt{h_1^2+x_2^2}/\lambda). \label{29}
\ee

In Fig. 6 we plot (29) for $T=1.070T_c$ and $R=2$ fm to observe  the
"double quasistring". The quasistring profile, $E_1(x_2)\equiv
\langle E_1(R/2,x_2) \rangle_{Q \bar Q}^{(b)}$ at $R\to\infty$, is
 \be
   E_1(x_2) = \pi B x_2^2 \exp(-|x_2|/\lambda).
\label{30}
\ee
The field in the centre of quasistring is absent.
The maximal value of the field is
\be
  E_1^{\max}=E_1(2\lambda)=1.7B\lambda^2.
\label{31}
\ee
In the coordinate space the quasistring resembles coaxial cable with
empty core and tube shell.
In Fig. 7 we plot  $\langle \veE(x_1,x_2) \rangle^{(b)}$ distribution
around $Q$ and $\bar Q$.

\section{Conclusions}

Results of our paper based on the analysis of the lattice data on correlation
functions at finite $T$ [1] give a full
support of the dynamical picture of deconfinement, which was first
suggested in [9].

Namely, confining and deconfining phases according to [9]
differ first of all in the vacuum fields, i.e., in the value of the
condensate and in the field correlators. It was argued in [9]
that color magnetic correlators and their contribution to the
condensate are kept intact across the temperature phase transition,
while the confining electric part abruptly disappears above
$T_c$. Both features are present in [1] and in the results of
the present paper. Indeed, one can see from Tables 5,6, that the
magnetic part of condensate is roughly constant around $T_c$. The
role of magnetic field above $T_c$ was mentioned repeatedly in the
literature, see recent lattice reviews [10], it reveals itself in
particular in creating nonzero spatial string tension \cite{9} and
so-called screening hadronic masses -- (see \cite{11}
and refs. therein).

The situation with electric fields is more subtle, as can be seen
from our results. In \cite{9} two possible situations have been
considered when nonperturbative part of $D^\mathrm{E}$ vanishes or stays
nonzero above $T_c$. From Figs. 4,6 one can clearly see
the field distributions in cases of two possible solutions, (a) and (b),
one with vanishing $D^\mathrm{E,NP}$,
another with nonvanishing but  oscillating $D^\mathrm{E,NP}$, both yielding
zero string tension in the deconfinement region.

In the case (a) the electric contribution to condensate is determined by
$D_1^\mathrm{E,NP}(0)$. While  $D^\mathrm{E,NP}(x)$ vanishes
identically,  the $D_1^\mathrm{E,NP}$ correlator is different from zero
above $T_c$   and its contribution to the condensate grows sharply with
temperature. Hence the role of $D_1^\mathrm{E,NP}$ appears in creating the
sharp rise of the "nonideality" ($\varepsilon-3p$) just above $T_c$ (cf.
Fig. 3).
One special remark is due to the regime (b), where $D^\mathrm{E,NP}(x)$ is
nonzero above $T_c$ but changing sign. This regime creates rather peculiar
picture of fields --- the "quasistring" with the empty core and surrounding
it tube shell at two correlation length distance from the quark axis (Fig.
6).

 As the regime (a) seems to be more natural from physical point
 of view, one should study in more detail the consequences of
 the strongly increasing with  $T$ $D_1^\mathrm{E,NP}(0)$ in the deconfining
region.

 The impossibility of resolving our present ambiguity (regimes (a)
 and (b)) calls for further numerical and analytical studies. It
 is necessary for understanding of the dynamics of the phase
 transition, where Polyakov loops and hence color-electric fields
 may play very important role.

The authors are grateful to A.Di Giacomo for useful remarks and
suggestions; the partial support of grants 00-02-17836 and 00-15-96786
 is gratefully acknowledged.

\clearpage
\bc
{\Large \bf List of tables}
\ec

\begin{table}[hbt]
\centering
\small
\caption{ Parameters of $D^B$.}
\vspace{0.3cm}
\label{table1}
 \begin{tabular}{|l|l|l|l|}
 \hline
  T   &  0.956$T_c$ & 0.978$T_c$ & 1.011$T_c$\\[5pt]
 \hline
  A, fm$^{-4}$ & 188.9$\pm$2.4 & 154$\pm$2 & 183.1$\pm$3.0\\
  $\lambda_A$, fm & 0.1917$\pm$0.0007 & 0.2047$\pm$0.0008 & 0.1852$\pm$0.0008\\
  B, fm$^{-4}$ & 7.7$\pm$0.6 & 7.7$\pm$0.8 & 1.87$\pm$0.13\\
  $\lambda_B$, fm & 0.380$\pm$0.008 & 0.344$\pm$0.007 & 1.11$\pm$0.04\\
  a & 1.11$\pm$0.06 & 1.46$\pm$0.07 & 0.64$\pm$0.03\\
  b & 0.34$\pm$0.03 & 0.47$\pm$0.04 & 0.35$\pm$0.02\\
  $\chi^2$/n & 0.62 & 1.28 & 1.43\\
 \hline
  T   &  1.034$T_c$ & 1.070$T_c$ & 1.131$T_c$\\[5pt]
 \hline
  A, fm$^{-4}$ & 128.8$\pm$2.3 & 111.9$\pm$2.0 & 150.9$\pm$2.5\\
  $\lambda_A$, fm & 0.210$\pm$0.001 & 0.2191$\pm$0.0012 & 0.2009$\pm$0.0010\\
  B, fm$^{-4}$ & 2.92$\pm$0.23 & 3.54$\pm$0.29 & 3.07$\pm$0.23\\
  $\lambda_B$, fm & 0.69$\pm$0.02 & 0.631$\pm$0.017 & 0.774$\pm$0.024\\
  a & 0.92$\pm$0.04 & 1.039$\pm$0.043 & 0.885$\pm$0.031\\
  b & 0.47$\pm$0.03 & 0.525$\pm$0.032 & 0.506$\pm$0.023\\
  $\chi^2$/n & 0.53 & 1.18 & 0.58\\
 \hline
 \end{tabular}
 \end{table}

\begin{table}[hbt]
\centering
\small
\caption{ Parameters of $D^E$ below $T_c$.}
\vspace{0.3cm}
\label{table2}
 \begin{tabular}{|l|l|l|}
 \hline
   T   &  0.956$T_c$ & 0.978$T_c$\\[5pt]
 \hline
  A, fm$^{-4}$ & 228$\pm$4 & 189.5$\pm$4.2\\
  $\lambda_A$, fm & 0.1823$\pm$0.0008 & 0.1812$\pm$0.0011\\
  B, fm$^{-4}$ & 10.8$\pm$0.5 & 14.3$\pm$0.6\\
  $\lambda_B$, fm & 0.435$\pm$0.011 & 0.411$\pm$0.006\\
  a & 3.1$\pm$0.4 & 0.99$\pm$0.08\\
  b & 0.9$\pm$0.2 & 0.71$\pm$0.18\\
  $\chi^2$/n & 0.39 & 1.7\\
 \hline
 \end{tabular}
 \end{table}

 \begin{table}[hbt]
\centering
\small
\caption{ Parameters of $D^E$ at $T=1.011T_c$.}
\vspace{0.3cm}
\label{table3}
 \begin{tabular}{|l|l|l|}
 \hline
     ~ &  (a): $D^{\mathrm{NP}}\equiv 0$ &
    (b): $D_{||}^{\mathrm{NP}}\equiv 0$\\[5pt]
 \hline
  B, fm$^{-4}$ & 2.46$\pm$0.24 & 1.96$\pm$0.34\\
  $\lambda_B$, fm & 0.682$\pm$0.016 & 1.072$\pm$0.047\\
  a & 1.74$\pm$0.05 & 1.36$\pm$0.04\\
  b & 0.86$\pm$0.04 & 0.68$\pm$0.03\\
  $\chi^2$/n & 1.7 & 1.05\\
 \hline
 \end{tabular}
 \end{table}

 \begin{table}[hbt]
\centering
\small
\caption{ Parameters of $D^E$ above $T_c$.}
\vspace{0.3cm}
\label{table4}
 \begin{tabular}{|l|l|l|l|}
 \hline
  T   &  1.034$T_c$ & 1.070$T_c$ & 1.131$T_c$\\[5pt]
 \hline
  B, fm$^{-4}$ & 80$\pm$19 & 235$\pm$53 & 519$\pm$99\\

  $\lambda_B$, fm & 0.121$\pm$0.05 & 0.105$\pm$0.004 & 0.37$\pm$0.03\\

  b & 0.63$\pm$0.03 & 0.48$\pm$0.04 & 0.41$\pm$0.03\\

  $\lambda_b$, fm & 0.86$\pm$0.05 & 1.0$\pm$0.1 & 0.86$\pm$0.08\\

  $\chi_1^2$ & 0.69 & 0.9 & 0.028\\[5pt]
 \hline
  (a)~ a & 1.21$\pm$0.04 & 0.94$\pm$0.04 & 0.86$\pm$0.03\\

  (a)~ $\chi_2^2$ & 1.26 & 0.9 & 2.9\\

  (b)~ a & 0.96$\pm$0.04 & 0.66$\pm$0.04 & 0.61$\pm$0.03\\

  (b)~ $\chi_2^2$ & 1.66 & 0.23 & 0.56\\
\hline
 \end{tabular}
 \end{table}

 \begin{table}[hbt]
\centering
\small
\caption{ Gluonic condensate below $T_c$.}
\vspace{0.3cm}
\label{table5}
 \begin{tabular}{|r|l|l|}
 \hline
  T   &  0.956$T_c$ & 0.978$T_c$\\[5pt]
 \hline
$D^\mathrm{B,NP}(0)+D_1^\mathrm{B,NP}(0), $ fm$^{-4}$  & 196.6$\pm$2.5 &
161.7$\pm$2.2\\
$D^\mathrm{E,NP}(0)+D_1^\mathrm{E,NP}(0), $ fm$^{-4}$ & 238.8$\pm$4.0 &
203.8$\pm$4.2\\[5pt]

$G_2(T)/G^0_2$ & 1.393$\pm$0.015 & 1.171$\pm$0.015\\[5pt]
 \hline
 \end{tabular}
 \end{table}

 \begin{table}[hbt]
\centering
\small
\caption{ Gluonic condensate above $T_c$.}
\vspace{0.3cm}
\label{table6}
 \begin{tabular}{|cr|l|l|l|l|}
 \hline
 ~ & T  & 1.011$T_c$&  1.034$T_c$ & 1.070$T_c$ & 1.131$T_c$\\[5pt]
 \hline
\multicolumn{2}{|r|}{$D^\mathrm{B,NP}(0)+D_1^\mathrm{B,NP}(0), $ fm$^{-4}$}
& 185$\pm$3 & 131.7$\pm$2.3 & 115.4$\pm$2.0 & 154.0$\pm$2.5\\[5pt]
$G_2(T)/G^0_2$ & (a)
 & 0.60$\pm$0.01 & 0.69$\pm$0.06 & 1.12$\pm$0.17 &2.16$\pm$0.32\\
~  & (b)  &0.59$\pm$0.01 & 0.422$\pm$0.007 & 0.370$\pm$0.006 &
0.493$\pm$0.008\\[5pt]
 \hline
 \end{tabular}
 \end{table}

\clearpage

\clearpage
\begin{figure}[t]
 \epsfxsize=12cm
  \centering
  \epsfbox{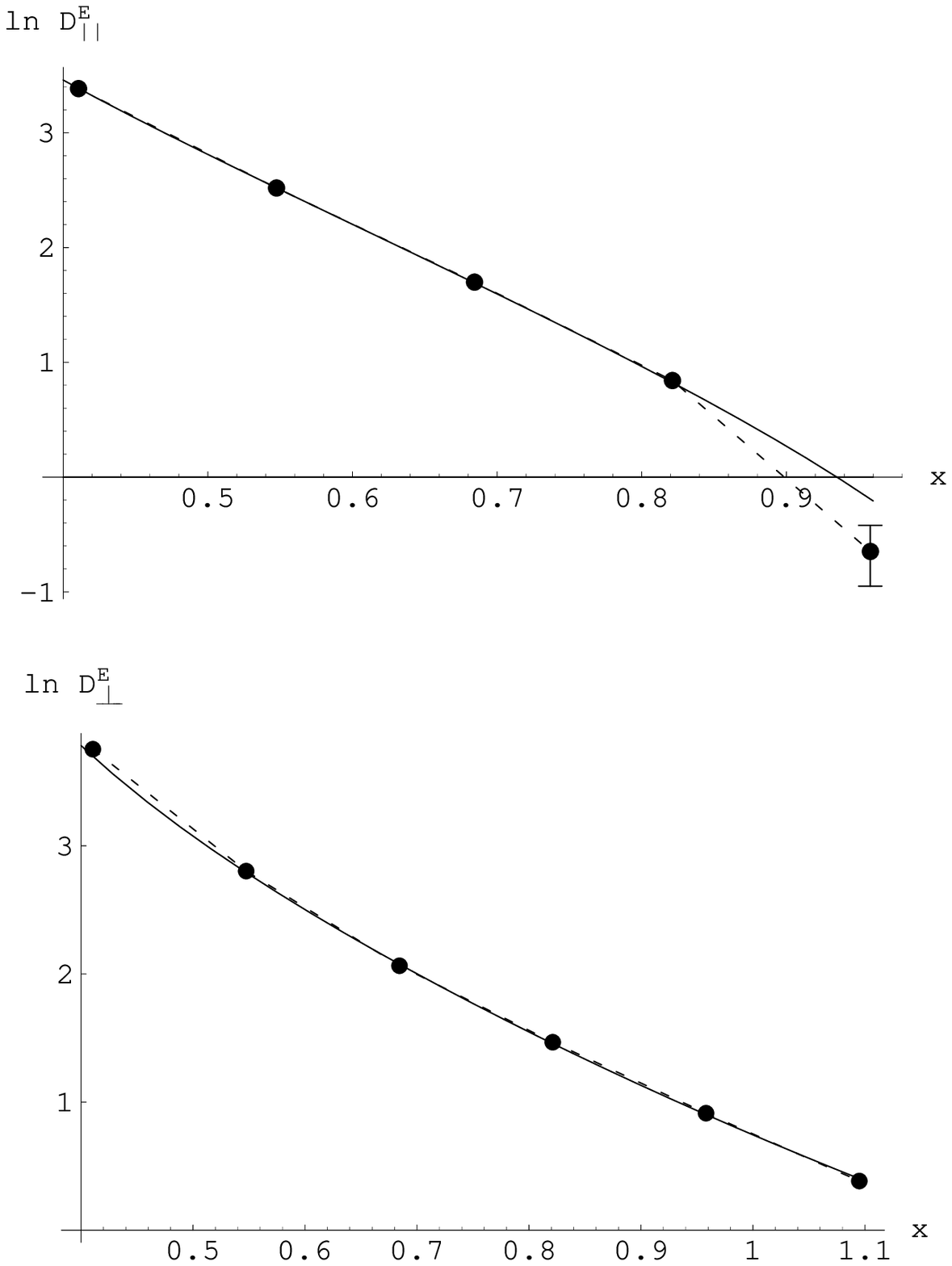}
   \caption{Fitted functions  $\ln D_{||}^\mathrm{E}(x)$  and $\ln
D_\bot^\mathrm{E}(x)$ (shown by solid lines) at $T=0.978T_c$.  $x$ is
measured in fm. $D_{||}^\mathrm{E}$  and $D_\bot^\mathrm{E}$ are measured in
fm$^{-4}$. Data are shown by points. Errors of data  are comparable with
size of the points.  The separately shown error  is two times enlarged
by hand for the improvement of the fit.}
\end{figure}

\clearpage

  \begin{figure}[!t]
 \epsfxsize=12cm
  \centering
  \epsfbox{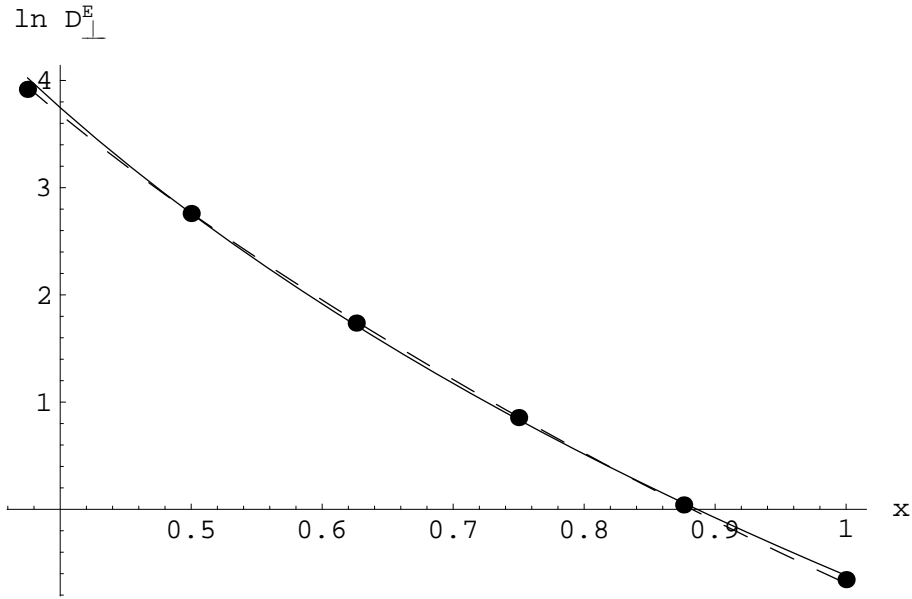}
   \caption{Fitted function   $\ln D_\bot^\mathrm{E}(x)$,
 shown by solid line in case (a) and dashed line in case (b) at
$T=1.070T_c$.    $x$ is measured in fm. $D_\bot^\mathrm{E}$ is measured in
fm$^{-4}$. Data are shown by points joined by dashed lines. Errors of data
are comparable with size of the points.}
\end{figure}

 \begin{figure}[t]
 \epsfxsize=12cm
  \centering
  \epsfbox{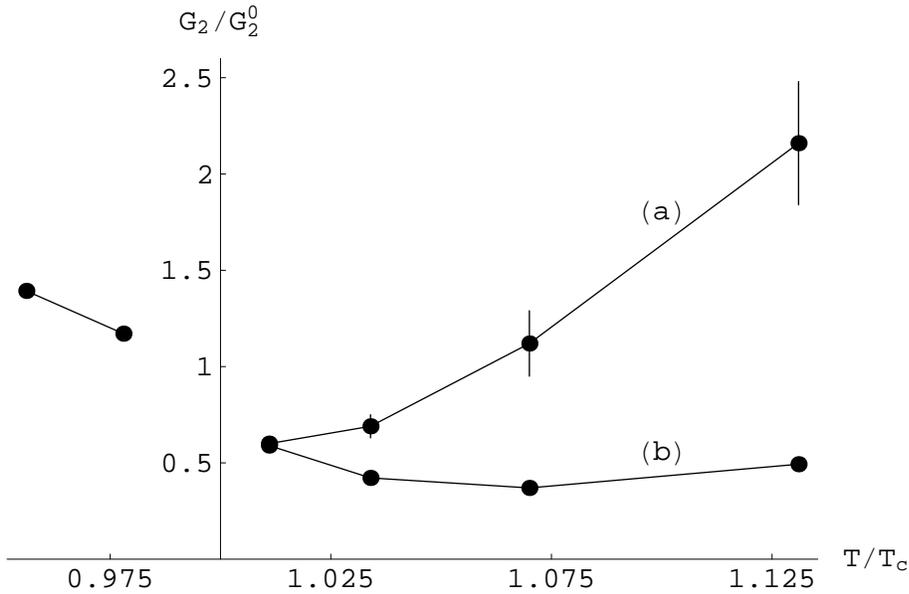}
   \caption{ Gluon condensate dependence on $T/T_c$ near the deconfinement
transition.  Condensate is measured in its zero temperature value
units. In the deconfinement region the magnetic part of condensate does not
considerably change. Electric fields give rapidly rising contribution  to
condensate in case (a) and do no contribute to condensate in case (b).}
\end{figure}

\clearpage

 \begin{figure}[t]
  \epsfxsize=16.5cm
   \centering
   \epsfbox{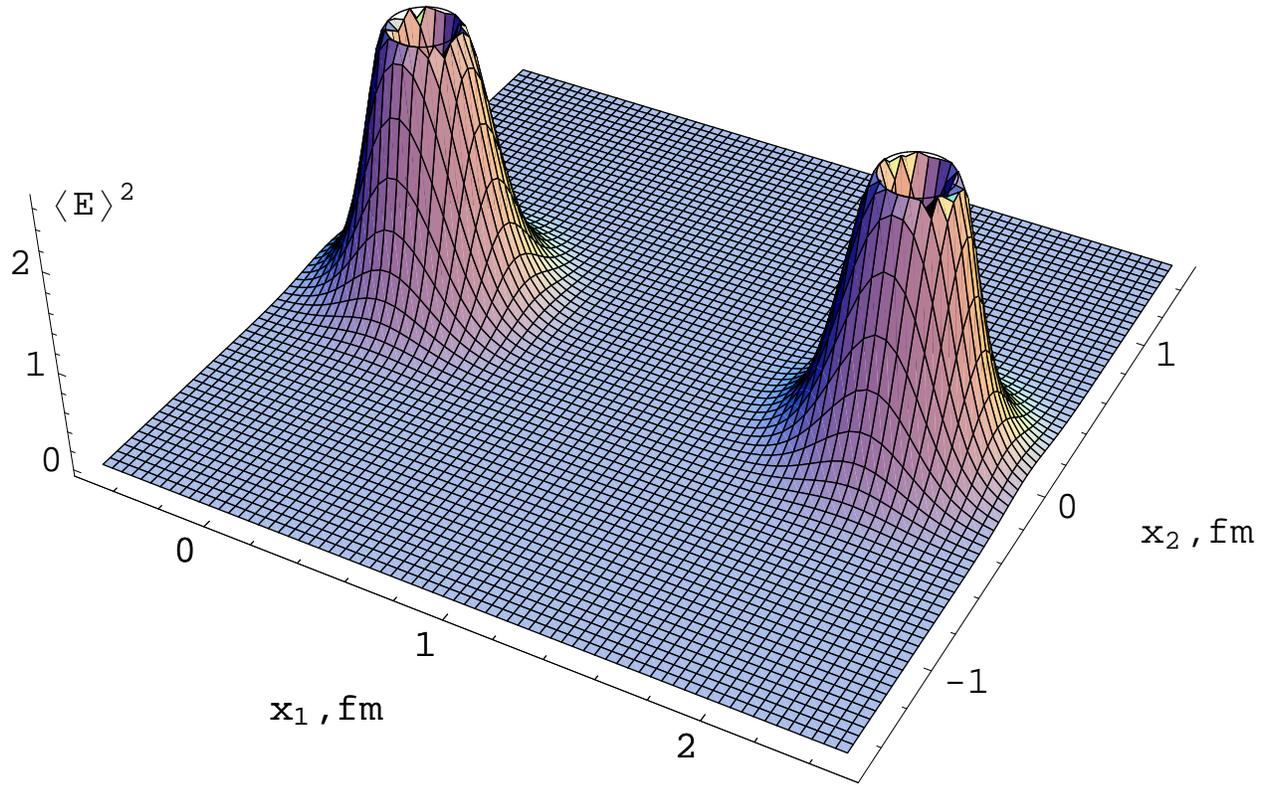}
   \caption{ The total field distribution $\langle \veE(x_1,x_2)\rangle^2$,
measured in fm$^{-4}$, in case (a) at $T=1.070T_c$. $x_1$ and $x_2$ are
measured in fm. $Q \bar Q$ separation $R=2$ fm.}
  \end{figure}

  \begin{figure}[t]
  \epsfxsize=8cm
   \centering
   \epsfbox{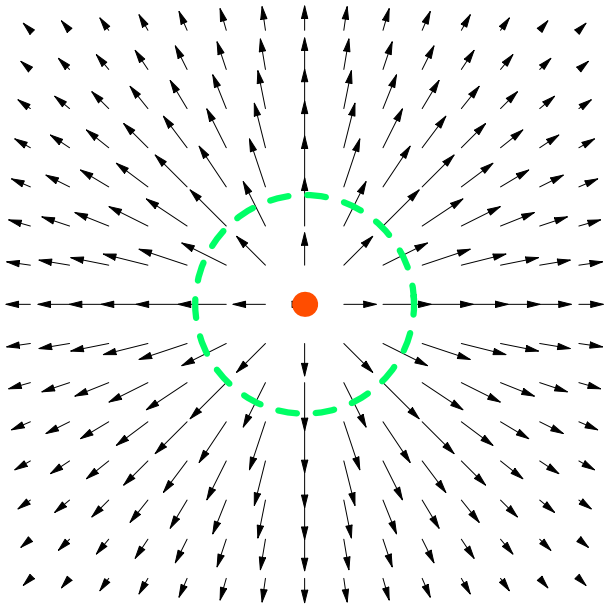}
   \caption{Vector distribution of $\langle \veE(x_1,x_2)\rangle$
in case (a). $-0.5$ fm$<x_1<0.5$ fm,~ $-0.5$ fm$<x_2<$0.5 fm.
Quark position  is marked with disk. Points of maximal value of field
are marked with dashed circle of radius $1.33\lambda$.}
  \end{figure}

\clearpage

  \begin{figure}[t]
  \epsfxsize=16.5cm
   \centering
   \epsfbox{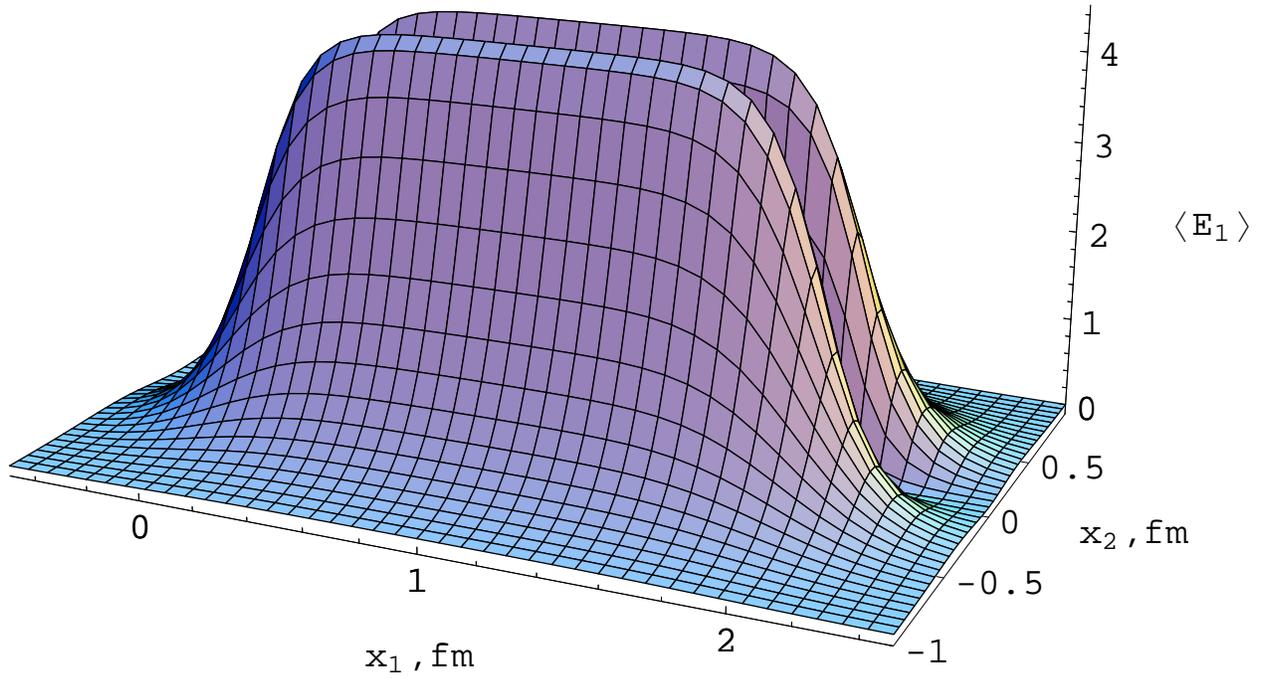}
   \caption{ Distribution of  $\langle E_1(x_1,x_2)\rangle$, measured in
fm$^{-2}$, in case (b) at $T=1.070T_c$.}
   \end{figure}

  \begin{figure}[t]
  \epsfxsize=11cm
   \centering
   \epsfbox{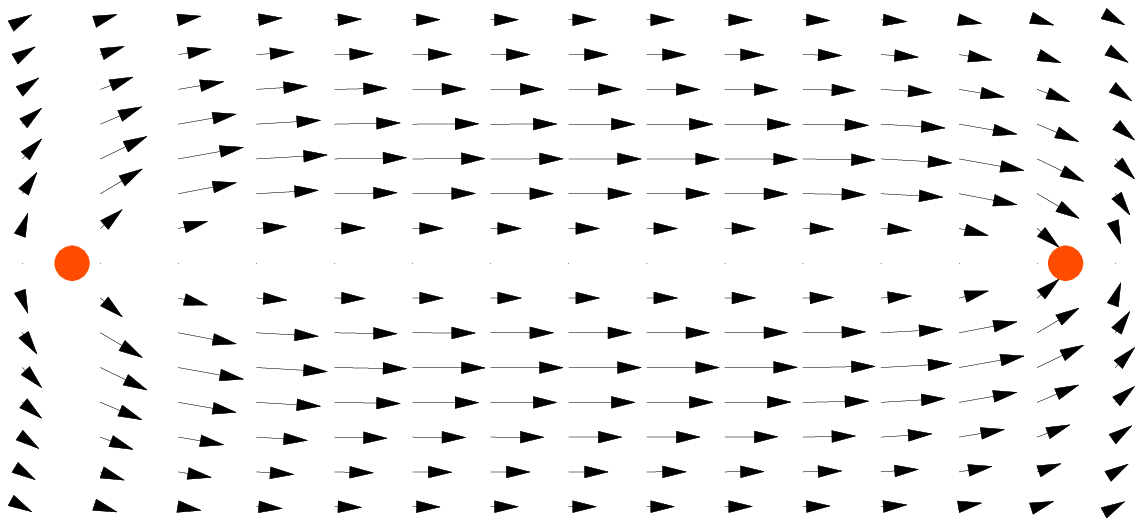}
\caption{ Vector distribution of $\langle \veE(x_1,x_2)\rangle$ in case
(b). Positions of $Q$ and $\bar Q$ are marked with
disks. $-0.1$ fm$<x_1<$2.1 fm, $-0.5$ fm$<x_2<$0.5 fm.}
 \end{figure}


\begin{thebibliography}{99}

   \bibitem {1}
     A.Di Giacomo, E.Meggiolaro, and H.Panagopoulos, Nucl.Phys. B {\bf 483}
(1997) 371


\bibitem{2}
 H.G.Dosch, V.I.Shevchenko, and Yu.A.Simonov, hep-ph/0007223.


 \bibitem {3}
 A.Di Giacomo, M.Maggiore, and S.Olejnik, Phys.Lett. B {\bf 236}, 199
 (1990); Nucl.Phys. B {\bf 347} 441 (1990)  \\
  L.Del Debbio, A.Di Giacomo, and Yu.A.Simonov, Phys.Lett. B {\bf 332}, 111
  (1994).


  \bibitem {4}
 H.G.Dosch, Phys.Lett. B {\bf 190}, 177  (1987);\\
 H.G.Dosch and Yu.A.Simonov, Phys.Lett. B {\bf 205}, 399 (1988);\\
  Yu.A.Simonov, Nucl.Phys. B {\bf 307}, 512  (1988).


\bibitem {5}
 Yu.A.Simonov,   Phys.Usp.   {\bf 39},  313 (1996).

\bibitem {6}
   Review of Particle Physics, Eur.Phys.J. C {\bf 15} (2000).

\bibitem {7}
    A.Di Giacomo, E.Meggiolaro, and H.Panagopoulos, preprint IFUP-TH 12/96,
    hep-lat/9603017.


 \bibitem {8}
 D.S.Kuzmenko and Yu.S.Simonov, Yad.Fiz. {\bf 64}, 110 (2001),
hep-ph/0010114.

\bibitem{9}
 Yu.A.Simonov, JETP Lett. {\bf 54}, 256 (1991);\\
 Yu.A.Simonov, JETP Lett. {\bf 55}, 605 (1992);\\
 Yu.A.Simonov, Yad.Fiz. {\bf 58}, 357 (1995);\\
 Yu.A.Simonov, Lectures at the E.Fermi International School, Varenna 1995,
 preprint ITEP-37-95.

\enlargethispage{2\baselineskip}

 \bibitem {10}
F.Karsch, Nucl.Phys.Proc.Suppl. {\bf 83}, 14 (2000);\\
S.Ejiri, preprint UTCCP-P-95, hep-lat/0011006.

\bibitem{11}
 E.L.Gubankova and Yu.A.Simonov, Phys.Lett. B {\bf 360}, 93 (1995).

\end{thebibliography}
\end{document}